\documentclass[conference]{IEEEtran}
\IEEEoverridecommandlockouts

\usepackage[ruled,linesnumbered]{algorithm2e}

\def\BibTeX{{\rm B\kern-.05em{\sc i\kern-.025em b}\kern-.08em
    T\kern-.1667em\lower.7ex\hbox{E}\kern-.125emX}}
    
\usepackage{CJKutf8}  

\usepackage{makecell}
\usepackage{xcolor}
\usepackage{pgfplots}
\usepackage{tikz}
\usepackage{comment}

\usepackage{filecontents}
\usepackage{threeparttable}

\usepackage{listings}
\usepackage{xcolor}

\lstdefinelanguage{Solidity}{
  keywords=[1]{pragma, contract, mapping, function, public, payable, view, returns, require, msg, address, uint256, bool},
  keywordstyle=[1]\color{blue}\bfseries,
  keywords=[2]{call},
  keywordstyle=[2]\color{purple}\bfseries,
  sensitive=true,
  comment=[l]{//},
  morecomment=[s]{/*}{*/},
  commentstyle=\color{green!50!black}\itshape,
  stringstyle=\color{orange},
  morestring=[b]"
}

\usepackage{pifont} %use the command \ding
\usepackage{amssymb}
\usepackage{bbding} %对勾叉号
\usepackage{pifont}

\usepackage{makecell}
\usepackage{colortbl} %tab color

\usepackage{fancyhdr} %add page number
\usepackage{hyperref}

\usepackage{multirow}
\usepackage{cite}
\usepackage{amsmath,amssymb,amsfonts}

\usepackage{textcomp}
\usepackage{xcolor}

\usepackage{subfigure}
\usepackage{pgfplots}
\usepackage{filecontents}

\usepackage{graphicx,subfigure}

\definecolor{bblue}{HTML}{4F81BD}
\definecolor{rred}{HTML}{C0504D}
\definecolor{ggreen}{HTML}{9BBB59}
\definecolor{ppurple}{HTML}{9F4C7C}

\usepackage{blindtext}
\usepackage{etoolbox,xstring,mfirstuc,textcase}
\usepackage{booktabs}
\usepackage{pgfplots}

\usepackage{tabularx,booktabs,makecell,multirow, caption}
\usepackage{enumitem} % to control Itemization spacing

\newcolumntype{L}{>{\arraybackslash}X}

\usepackage{xcolor}
\usepackage{tikz}
\usepackage{pgfplots}

\usepackage{authblk}

\definecolor{findOptimalPartition}{HTML}{D7191C}
\definecolor{storeClusterComponent}{HTML}{FDAE61}
\definecolor{dbscan}{HTML}{ABDDA4}
\definecolor{constructCluster}{HTML}{2B83BA}

\usepackage[utf8]{inputenc}
\usepackage{listings}
\begin{document}
%=================================================   
\title{Logic Meets Magic: LLMs Cracking Smart Contract Vulnerabilities}
%=================================================   
%=================================================   

%\author{Anonymous}
%\begin{comment}
\author{ZeKe Xiao$^1$, Qin Wang$^{1,2}$, Hammond Pearce$^1$, Shiping Chen$^{1,2}$}
\affil{$^1$\textit{UNSW Sydney} $|$ $^2$\textit{CSIRO Data61}, Australia}
%\end{comment}

%=================================================    
\maketitle
%=================================================  

\begin{abstract}

Smart contract vulnerabilities caused significant economic losses in blockchain applications. Large Language Models (LLMs) provide new possibilities for addressing this time-consuming task. However, state-of-the-art LLM-based detection solutions are often plagued by high false-positive rates.

In this paper, we push the boundaries of existing research in two key ways. First, our evaluation is based on Solidity v0.8, offering the most up-to-date insights compared to prior studies that focus on older versions (v0.4). Second, we leverage the latest five LLM models (across companies), ensuring comprehensive coverage across the most advanced capabilities in the field.

We conducted a series of rigorous evaluations. Our experiments demonstrate that a well-designed prompt can reduce the false-positive rate by over 60\%. Surprisingly, we also discovered that the recall rate for detecting some specific vulnerabilities in Solidity v0.8 has dropped to just 13\% compared to earlier versions (i.e., v0.4). Further analysis reveals the root cause of this decline: the reliance of LLMs on identifying changes in newly introduced libraries and frameworks during detection.

%The rapid development of blockchain applications has also led to significant economic losses due to vulnerabilities in their core component--smart contracts. Meanwhile, rising abruptly in recent years, LLMs offer new insights for overcoming this time-consuming and challenging task.\par
%However, current studies and research are quite limited, commonly revealing a high false positive rate in LLM detection.\par
%Based on this, we first evaluated the performance of current mainstream LLMs, particularly their false positive rates, and discussed how to reduce false positive reports through reasonable prompt design. Our experiment reveals that a well-designed prompt can reduce the false positive rate by more than 60\%. \par Furthermore, we are the first to conduct experiments on mainstream LLMs for detecting vulnerabilities in smart contracts written in Solidity version 0.8. Compared to successfully detecting the same vulnerabilities in Solidity version 0.4, the recall rate of LLMs may drop to 13\%. We analyzed the experimental results and revealed the root causes of the new challenges LLMs face in identifying vulnerabilities, such as their reliance on checking new libraries during detection.    

\end{abstract}

\smallskip

%=================================================   
\section{Introduction}
%=================================================  

A smart contract is an automatable and enforceable agreement. It is automatable by computer, though some aspects may require human input and control, and enforceable either through legal mechanisms or tamper-proof execution of code~\cite{clack2016smart}. Smart contracts are a fundamental component of blockchain technology~\cite{sarmah2018understanding}.
Unlike traditional contracts, smart contracts automate agreement execution in a distributed environment when predefined conditions are met~\cite{khan2021blockchain}. Their efficiency and reliability have led to widespread adoption in blockchain systems, enabling the development of Web3~\cite{wang2022exploring}, including decentralized applications (DApps), Decentralized Finance (DeFi)~\cite{werner2022sok, jiang2023decentralized}, Non-Fungible Tokens (NFTs)\cite{wang2021non}, and Game Finance (GameFi)~\cite{proelss2023gamefi}.

Smart contract-powered on-chain DApps and protocols have gained immense popularity in recent years, with Bitcoin and Ethereum ETPs collectively holding US\$65b in on-chain assets as of 2024~\cite{W1}. However, this growth has also exposed vulnerabilities in smart contracts, leading to significant financial losses. According to a SlowMist~\cite{sw}, in August 2024 alone, these vulnerabilities resulted in losses exceeding US\$316m.

Manually auditing smart contracts is labor-intensive~\cite{wang2023evaluation}, and current smart contract scanners (over 17 types) have demonstrated poor performance~\cite{sendner2023smarter}. While large language models (LLMs) offer new opportunities for tackling this challenge, having shown high efficiency in program analysis~\cite{pearce2023examining, thapa2022transformer} and code generation~\cite{pearce2022asleep}, their potential for smart contract vulnerability detection remains underexplored.

Existing evaluations of LLMs for this purpose are limited (evidence in \textbf{Table~\ref{tab-stateofart}}), often focusing on the GPT series and revealing high false positive rates (FPR)~\cite{david2023you, chen2023chatgpt}. Moreover, these studies primarily use datasets written in outdated Solidity versions (v0.4), neglecting the nuances of detection in Solidity v0.8, the more widely used version. Additionally, most detection methods rely on a single LLM (e.g., GPT-3.5 or GPT-4), restricting their cross-platform applicability.

\smallskip
To address these challenges, we posed the following research questions (short for \textbf{RQ}s):

\begin{itemize}
    \item \textbf{RQ.1:} What is the FPRs of mainstream LLMs when analyzing a dataset of non-vulnerable top contracts?
    \item \textbf{RQ.2:} Can we decrease false positive rate through reasonably prompt design?
    \item \textbf{RQ.3:} What are the limitations of general LLMs in detecting vulnerabilities in smart contracts written in Solidity v0.8? Any differences to v0.4?
\end{itemize}

Our study centers on investigating the potential limitations of LLMs in identifying vulnerabilities and examining possible measures to enhance their performance.

\begin{itemize}
    \item  We conducted an experiment on the TOP200 dataset to evaluate the basic performance of five up-to-date LLMs in detecting vulnerabilities, particularly focusing on the false positive rate. After the pre-test, the Max Tokens for each LLM were sufficient to allocate general smart contracts, eliminating the need for division. Then, we successfully reduced the false positive rate through a reasonably prompt design and validated it with an experiment. This new design can help researchers determine the lower bound of LLM's false positive rate.
    \smallskip
   \item We conducted another experiment on the Web3Bugs dataset to evaluate the performance of evaluated LLMs in identifying  different types of vulnerabilities, including re-entrancy, arithmetic issues, denial of service, access control, manipulated price, and oracle issues. This marks the first evaluation of the performance of LLMs using smart contracts written in Solidity v0.8, which is currently the most widely utilized version. In contrast, previous studies were based on Solidity v0.4 and involved the analysis of much shorter and simpler codes.

   \smallskip
   \item Furthermore, we observed differences in the detection abilities of general LLMs between Solidity v0.8 and v0.4. By analyzing the experimental results, we uncovered three important limitations and root causes of LLMs' detection performance in Solidity v0.8. In the evaluation, LLMs heavily rely on recognizing established libraries and frameworks, which often leads to the oversight of other potential issues, particularly Arithmetic vulnerabilities and Re-entrancy vulnerabilities. Moreover, higher detection capabilities often result in a higher false positive rate. These findings offer novel perspectives
and strategies for improving LLMs’ performance.
   \end{itemize}

%================================================
\section{Technical Warmups}
\label{sec-tech}
%================================================

\subsection{Core Concepts}

\noindent\textbf{Blockchain.}
A blockchain is a distributed ledger with growing lists of records (blocks) that are securely linked together via cryptographic hashes\cite{BC}. It was first proposed by S. Nakamoto\cite{nakamoto2008bitcoin}, a peer-to-peer network that sits on top of the internet, and was introduced to public in October 2008 as part of a proposal for bitcoin, a virtual currency system that eschewed a central authority for issuing currency, transferring ownership, and confirm transactions. Bitcoin is the first blockchain application\cite{iansiti2017truth}. Additionally, Blockchain has numerous benefits such as decentralization, persistency, anonymity and audit-ability. There is a wide spectrum of blockchain applications ranging from cryptocurrency, financial services, risk management, internet of things (IoT) to public and social services\cite{zheng2018blockchain}.

\smallskip
\noindent\textbf{Smart Contract.}
A smart contract is an intelligent agent. In other words, it is a computer program capable of making decisions when certain preconditions are met. The intelligence of an agent depends on the complexity of a transaction it is programmed to perform\cite{kolvart2016smart}. In a smart contract, contract clauses written in computer programs will be automatically executed when predefined conditions are met. Smart contracts consisting of transactions are essentially stored, replicated, and updated in distributed blockchains~\cite{zheng2020overview}.

\smallskip
\noindent\textbf{Large Language Models.}
Typically, LLMs refer to Transformer
language models that contain hundreds of billions (or more) of parameters, which are trained on massive text data. LLMs exhibit strong capacities to understand natural language and solve complex tasks (via text generation)\cite{zhao2023survey}.

With the increase in capabilities, LLMs are able to autonomously exploit one-day vulnerabilities in real-world systems. When given the CVE description, GPT-4 is capable of exploiting 87\% of these vulnerabilities\cite{fang2024llm}. Additionally, LLMs can achieve end-to-end program repair~\cite{jin2023inferfix}.

\subsection{Key Questions Before Main-dish}

\noindent\textbf{What are smart contract vulnerabilities?}
Similar to any other software, smart contracts are susceptible to bugs and vulnerabilities. Given that smart contracts
are directly associated with cryptocurrencies, the potential
financial losses resulting from undiscovered vulnerabilities
can be significant\cite{sendner2024large}. According to a SlowMist report~\cite{Sl}, in November 2024 alone, vulnerabilities in smart contracts have resulted in financial losses exceeding \$9.38 million. 

Common vulnerabilities include Reentrancy, Timestamp Dependence, Denial of Service, Oracle Issues, arithmetic Issues, Access Control, Price Manipulation~\cite{sayeed2020smart}. 

Here, we offer an example for Reentrancy vulnerablities. 

\begin{lstlisting}[caption={Vulnerable Solidity Contract}, label={lst:reentrancy}]
contract VulnerableContract {
    mapping(address => uint256) public balances;
    function deposit() public payable {
        balances[msg.sender] += msg.value;   }
    function withdraw(uint256 _amount) public {
        require(balances[msg.sender] >= _amount, "Insufficient balance");
        (bool success, ) = msg.sender.call{value: _amount}("");
        require(success, "Transfer failed");
        // Update the balance AFTER the transfer (vulnerable)
        balances[msg.sender] -= _amount; }
    function getBalance() public view returns (uint256) {
        return address(this).balance;  }
}
\end{lstlisting}

The withdraw function transfers Ether to the caller using an external $\mathsf{call}$ before updating the user's balance. This allows a malicious contract to re-enter the withdraw function during the external call and withdraw funds repeatedly before the balance is updated, draining the contract's funds. The vulnerability arises from improper ordering of operations, specifically failing to update the state before interacting with external accounts. To prevent this, developers should follow the ``checks-effects-interactions" pattern, update the state before making external calls, or use reentrancy guards to block recursive calls.

\smallskip
\noindent\textbf{Why can LLM help to detect those vulnerabilities?}
LLMs are effective in detecting smart contract vulnerabilities, such as reentrancy, due to their advanced understanding of programming semantics and patterns. Trained on extensive code datasets, LLMs can analyze contracts, identify risky patterns like external calls before state updates, and flag vulnerabilities linked to known attack vectors. Their contextual understanding allows them to evaluate the logic and flow of a contract, uncovering subtle issues often missed by static analysis tools.

Additionally, LLMs offer unmatched scalability and efficiency, analyzing large volumes of smart contracts in real-time, which is essential for blockchain ecosystems with frequent deployments. By automating insights, highlighting risks, and suggesting fixes, LLMs complement manual audits and help secure increasingly complex contracts, such as those written in Solidity v0.8, making them a critical tool for mitigating risks.

%================================================
\section{Methodology}
\label{sec-metho}
%================================================

\begin{figure}[!h]
\centering
\includegraphics[width=\linewidth]{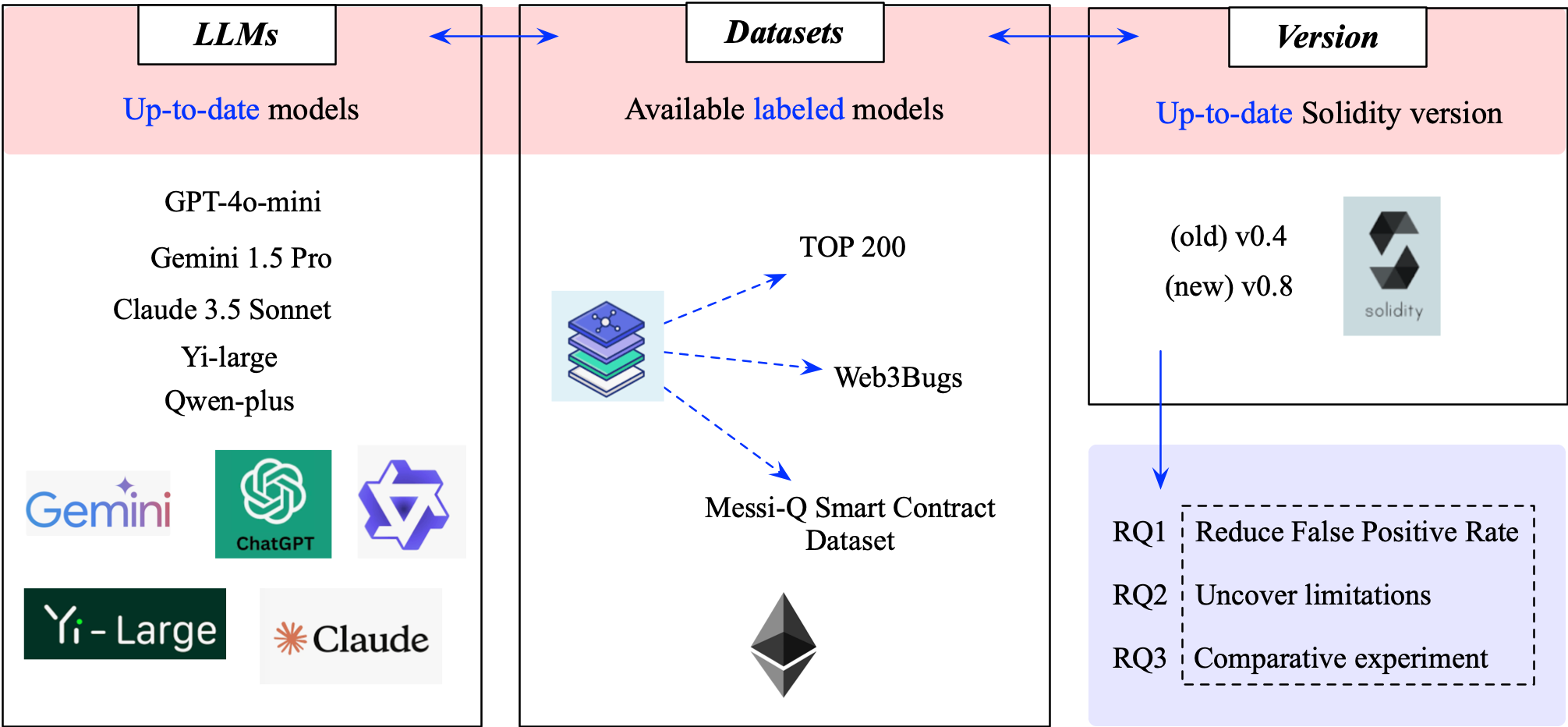}
\caption{Methodology Overview}
\label{22}
\end{figure}

%\begin{figure}[!h]
%\centering
%\includegraphics[scale=0.2]{22.png}
%\caption{Research Procedure}
%\label{22}
%\end{figure}

\subsection{Overview}

\textbf{Fig.~\ref{22}} illustrates the research procedure. Our study integrates five up-to-date LLMs (GPT-4o-mini, Gemini 1.5 Pro, Claude 3.5 Sonnet, Yi-large, and Qwen-plus) and assesses their performance across three curated datasets: the TOP 200 dataset, the Web3Bugs dataset, and the Messi-Q Smart Contract Dataset. To ensure relevance, our experiments target Solidity v0.8, a widely used version, addressing gaps in prior studies that focused on outdated versions like v0.4 (also performing evaluation). Our research process is structured around three research questions (RQs): analyzing false positive rates (RQ1), designing effective prompts to mitigate these rates (RQ2), and uncovering LLM limitations in vulnerability detection (RQ3).
We detail each items below.

\subsection{Dataset}
\subsubsection{\underline{TOP200}}
The first dataset used in the experiment is TOP200, including 303 open-source contract projects and approximately 1000 smart contracts from six mainstream Ethereum-compatible chains\cite{yi2022blockscope}. Since these projects are well-audited and widely used, it is reasonable to assume that they do not contain notable vulnerabilities\cite{sun2024gptscan}. This dataset allows for a comprehensive analysis, as it includes projects written in both Solidity v0.4 and v0.8.

\subsubsection{\underline{Web3Bugs}}
The second dataset used in the experiment is Web3Bugs, which provides a thorough analysis of exploitable bugs extracted from code4rena and classify each bug according to its nature\cite{zhang2023demystifying}.\par 
In our experiment, each experimental group was assigned approximately 10 projects from Web3Bugs, with each project containing an indeterminate number of smart contracts, some of which were labeled with specific types of vulnerabilities.\par This dataset was used for the following key reasons:\par a. Each smart contract experienced full detailed analysis of professional auditors and Blockchain companies; \par
b. Most of the smart contracts were written in Solidity v0.8, which is prevalently used at this stage;\par c. The vulnerabilities labeled in this dataset cover common faults and fulfill our experimental requirements.

\subsubsection{\underline{M.Q-Smart-Contract-Dataset}}
The third dataset used in the experiment is M.Q-Smart-Contract-Dataset\cite{liu2023rethinking}, containing data labeled with eight types of vulnerabilities respectively, including block number dependency (BN), dangerous delegatecall (DE), dangerous delegatecall (DE), ether strict equality (SE), reentrancy (RE), timestamp dependency (TP), unchecked external call (UC), and arithmetic issues (AI).\par
The smart contracts in this dataset were all written in Solidity v0.4 and were used in our comparative experiment to provide a comparison in detail.

\subsection{LLMs}

In order to evaluate the general performance of mainstream LLMs and compare their attributes , we collected five types of LLMs from different platforms, including GPT-4o-mini, Gemini 1.5 pro, Claude 3.5 Sonnet, Yi-large, and Qwen-plus. The details are shown in the \textbf{Table~\ref{LLMs}}.

\begin{table}[h]
    \centering
    \caption{LLMs in experiment}    
    \begin{tabular}{c|cccc}
        \toprule
        \multicolumn{1}{c}{\textbf{Model API} } & \textbf{Params} & \textbf{Max Tokens} & \textbf{\makecell{Knowledge \\ Cut-off}} \\ 
        \midrule
        GPT-4o-mini
         & UNKNOWN & 128K & 10/2023 \\ 
        Gemini 1.5 pro & 175B & 2097K & 11/2023 \\ 
        Claude 3.5 Sonnet
        & UNKNOWN & 200K & 04/2024 \\ 
        Yi-large & 100B & 32K & UNKNOWN \\ 
        Qwen-plus & UNKNOWN & 131K & UNKNOWN \\ 
        \bottomrule
    \end{tabular}
    \label{LLMs}
\end{table}

\subsubsection{\underline{GPT-4o-mini}}

GPT-4o-mini surpasses GPT-3.5 Turbo and other small models on academic benchmarks across both textual intelligence and multimodal reasoning, and supports the same range of languages as GPT-4o\cite{W3}.
It excels in mathematical reasoning and coding tasks, outperforming previous small models on the market. On MGSM, measuring math reasoning, GPT-4o mini scored 87.0\%, compared to 75.5\% for Gemini Flash and 71.7\% for Claude Haiku. GPT-4o mini scored 87.2\% on HumanEval, which measures coding performance, compared to 71.5\% for Gemini Flash and 75.9\% for Claude Haiku.

\subsubsection{\underline{Gemini 1.5 pro}}

Gemini 1.5 Pro is trained by Google Deep Mind. It achieves near-perfect recall on long-context retrieval tasks across modalities, unlocking the ability to accurately process large-scale documents, thousands of lines of code, hours of audio, video, and more\cite{W4}. It is able to reason across 100000 lines of code and give helpful explanations\cite{team2023gemini}.

\subsubsection{\underline{Claude 3.5 Sonnet}}

Claude 3.5 Sonnet can independently write, edit, and execute code with sophisticated reasoning and troubleshooting capabilities. It handles code translations with ease, making it particularly effective for updating legacy applications and migrating code bases\cite{W5}.

\subsubsection{\underline{Yi-large}}
Yi-Large is a model for generating code as well as logic and mathematical reasoning. It is the latest proprietary dense model of the Yi Series State of the Art LLM from 01.AI. The model was trained with significant improvement from
the November 2023 Yi-34B open-source model detailed in this tech report. The larger and enhanced Yi-Large model demonstrates exceptional performance on all the benchmarks, especially code, math, and comprehensive reasoning. Overall, Yi-Large performs on par with GPT-4 and Claude3\cite{W6}.

\subsubsection{\underline{Qwen-plus}}
QWEN is a comprehensive language model series that encompasses distinct models with varying parameter counts. It includes QWEN, the base pretrained language models, QWEN-CHAT, and coding-specialized models, CODE-QWEN and CODE-QWEN-CHAT, as well as mathematics-focused models. 
Qwen2 surpasses most prior open-weight models, including its predecessor Qwen1.5, and exhibits competitive performance relative to proprietary models across diverse benchmarks on language understanding, generation, multilingual proficiency, coding, mathematics, and reasoning. The flagship model, Qwen2-72B, showcases remarkable performance: 84.2 on MMLU, 37.9 on GPQA, 64.6 on HumanEval, 89.5 on GSM8K, and 82.4 on BBH as a base language model. The instruction-tuned variant, Qwen2-72B-Instruct, attains 9.1 on MT-Bench, 48.1 on Arena-Hard, and 35.7 on LiveCodeBench \cite{bai2023qwen}\cite{yang2024qwen2}.

%================================================
\section{Detection Results}
\label{sec-results}
%================================================
\subsection{RQ1: Measuring False Positive Rate of mainstream LLMs}

\subsubsection{\underline{Evaluation}}

In RQ1, aiming to measure the false positive rate of mainstream LLMs in analyzing smart contracts, we collected smart contracts from the TOP200 datasets and scanned each Solidity document using LLMs. Although some of the documents are stable library files or system documents, each LLM scanned the same SOL documents, so this would not influence the experimental results.
Along with the experimental results, six metrics were used to evaluate the performance of LLM, including TN, TP, FP, FN, Accuracy, and False Positive Rate. Among them, Accuracy and False Positive Rate are derived from the other metrics.

\begin{itemize}
    \item \textbf{TP} is the number of True Positives. One true positive is counted when LLM successfully detects a ground-truth vulnerability after scanning a smart contract of dataset. 

    \item \textbf{TN} is the number of True Negatives. One true negative is counted when LLM correctly does not report any vulnerability after scanning a smart contract of dataset.

    \item \textbf{FP} is the number of False Positives. One False positive is counted when LLM incorrectly reports oneor more vulnerabilities after scanning a smart contract of dataset.

    \item \textbf{FN} is the number of False Negatives. One false negative is counted when LLM fails to detect the ground-truth vulnerability after scanning a smart contract of dataset.

    \item \textbf{False Positive Rate} (FPR) indicates a given condition exists when it does not. \par      

\end{itemize}

\begin{equation}
    FPR = \frac{FP}{FP+TN}
    \label{eq:division_formula}
\end{equation}

In the evaluation, we established a dialogue with the model through the API. The prompt is first sent to the model to initiate the evaluation. Then, the model reads the code of each smart contract and inputs it into the model.

The prompt was designed as:
\begin{center}
 %  \fbox{
   \colorbox{blue!10}{
         \begin{minipage}{0.91\linewidth}
            You are a smart contract auditor. ~\\
            Please review the following smart contract in detail carefully. Is the following smart contract vulnerable to any attacks? Please only answer yes or no. If yes, please answer with one main vulnerability. [\textbf{content}] 
         \end{minipage}
      }
  % }
\end{center}

\smallskip
In the prompt, ``\textbf{content}" is the code of each smart contract. Additionally, the info of models has been recorded in \textbf{Table.~\ref{LLMs}}. All of them are the latest models of their respective series by the time of writing the paper. Among them, Gemini has released 2.0 Pro version, but it is not stable enough to be used in evaluation. The temperature parameter of  for each model was set at 0.7, a default value, to ensure fostering innovation and stability during answer generation, while other parameters were also kept at their default settings.

\smallskip
\subsubsection{\underline{Result}}

The evaluation results of these models are listed in \textbf{Table~\ref{fal_exp}}, which shows that the false positive rates of all the models in the experiment are generally high. Among them, the false positive rates of GPT-4o-mini and Claude 3.5 Sonnect reached 0.85 and 0.78, respectively, while Yi-large and Qwen-plus demonstrated stable performance, maintaining a false positive rate around 0.5.

\begin{table}[h]
    \centering
    \caption{False Positive Rate in RQ1}    
    \begin{tabular}{c|cccc}
        \toprule
        \multicolumn{1}{c}{\textbf{Model API} }& \textbf{TP} & \textbf{FP} & \textbf{TN} & \textbf{\makecell{FPR}} \\ 
        \midrule
        GPT-4o-mini
         & 0 & 846 & 151 & 0.85 \\ 
        Gemini 1.5 pro & 0 & 706 & 291 & 0.71 \\ 
        Claude 3.5 Sonnet
        & 0 & 777 & 220 & 0.78 \\ 
        Yi-large & 0 & 447 & 550 & 0.45 \\ 
        Qwen-plus & 0 & 522 & 475 & 0.52 \\ 
        \bottomrule
    \end{tabular}    
    \label{fal_exp}
\end{table}

\begin{center}
\fbox{%
%    \colorbox{teal!15}{
     \begin{minipage}{0.9\linewidth}
    \textbf{Answer to RQ1.} \par
     The false positive rate of current mainstream LLMs in vulnerability detection is generally higher than 50\%.
      \end{minipage}
% }
}
\end{center}

\subsection{RQ2: Decreasing FPR via Reasonable Prompt?}

\subsubsection{\underline{Evaluation}}

A high false positive rate is a common attribute reported in related experiments on detecting vulnerabilities in smart contracts or programs with LLMs. In RQ1, we found that this issue occurs in general LLMs. In response to the extremely high false positive rate, we further explore a method to reduce the false positive rate of these models, and evaluate the potential lower bound of their false positive rate.\par
To achieve this reduction without fine-tuning or introducing other methods, the first step is redesigning the prompt.\par
After checking the prompt of RQ1, we found that RQ1's prompt did not clearly specify what vulnerabilities the models were expected to detect. Even though a binary classification would have been simple enough to understand, there was no clear direction for detecting vulnerabilities. As a result, the LLMs made more mistakes.\par 
Based on this evaluation, we proposed the following hypothesis: LLMs would have a lower false positive rate when detecting a specific vulnerability. 

Hence, we redesigned the prompt:

\begin{center}
  % \fbox{
   \colorbox{blue!10}{
         \begin{minipage}{0.89\linewidth}
            You are a smart contract auditor. ~\\
            Please review the following smart contract in detail carefully.\par Please confirm whether the smart contract project has “\textbf{A}” vulnerability, and only answer yes or no.[\textbf{content}] 
         \end{minipage}
      }
 %  }
\end{center}

\smallskip
In the actual prompt, ``\textbf{A}" represents the types of vulnerabilities and was replaced with the specific name of vulnerability, such as ``Access Control", while ``\textbf{content}" refers to the code of each smart contract.

For a detailed experiment, we used the dataset Web3bugs. 
In this experiment, we selected four types of vulnerabilities from the official DASP10 website\cite{W2}, while two other types were adopted from other cases\cite{Or}. The vulnerabilities include \textit{re-entrancy} (RE), 
\textit{arithmetic issues} (AI), \textit{denial of service} (DoS), \textit{access control} (AC),   
\textit{manipulated price} (MP), and \textit{oracle issues} (OI). DASP10 lists the top 10 vulnerabilities in smart contracts, but since it was established in 2018, some new vulnerabilities may not be included, and some classical vulnerabilities may have already been patched.

The vulnerabilities we studied were listed in \textbf{Table~\ref{Vul}}.

\begin{table}[!h]
    \centering
    \caption{Description of vulnerabilities}    
    \begin{tabular}{c|l}
        \toprule
        \multicolumn{1}{c}{\textbf{Vulnerability}}  & \multicolumn{1}{c}{\textbf{Description} }\\ 
        \midrule
        Re-entrancy  & Make a recursive call back to some functions. \\ 
        Arithmetic issues & Over/underflows and wrong calculation.   \\ 
        DoS  & Congestion of time-consuming computations.\\
        Access control   & The user without permission can execute functions. \\ 
        Manipulated price   & Tricking in using an incorrect valuation for a token.  \\
        Oracle issues  & The security about third-party oracles. \\ 
        \bottomrule
    \end{tabular}
    \label{Vul}
\end{table}

With the inclusion of more data types, additional metrics were introduced to evaluate the results, including Recall Rate, Accuracy, and Precision.\par

\begin{itemize}
    \item  \textbf{Recall Rate} evaluates the percentage of relevant items identified by the model.\par

\begin{equation}
    Recall Rate = \frac{TP}{TP+FN}
    \label{eq:division_formula}
\end{equation}

    \item \textbf{Accuracy} evaluates how close a given set of measurements are to their true value.\par

\begin{equation}
    Accuracy = \frac{TP+TN}{TP+FP+TN+FN}
    \label{eq:division_formula}
\end{equation}

    \item \textbf{Precision} evaluates how close the measurements are to each other.\par

\begin{equation}
    Precision = \frac{TP}{TP+FP}
    \label{eq:division_formula}
\end{equation}

\end{itemize}

\smallskip
The experimental result is recorded in \textbf{Table~\ref{tab:result}}.\par
During testing, GPT-4o-mini consistently answered with ``\textbf{YES}" or ``\textbf{NO}", a binary classification, while some other models generated additional  text like
\smallskip

\begin{center} 
   %\fbox{
   \colorbox{red!15}{
         \begin{minipage}{0.89\linewidth}
            \textbf{``Sorry, I can't evaluate whether this smart contract has Access Control vulnerability"}. 
         \end{minipage}
      }
 %  }
\end{center}
\smallskip

In our evaluation, cases with no specific vulnerability are categorized as ``\textbf{TN}", True Negative. On the other hand, if a real vulnerability is present, the case would be recorded as ``\textbf{FN}", that is False Negative.\par

\smallskip
\subsubsection{\underline{Result}}

The results indicate a significant decrease in the false positive rate of these models. A comparison of the models is shown in \textbf{Fig.~\ref{fig1}}. The main factor contributing to this improvement is that, in RQ2 experiment, the models were tasked with evaluating the presence of only one primary vulnerability, which allowed them to follow on a single direction.\par
\smallskip

\begin{center}
\fbox{%
    
    \begin{minipage}{0.9\linewidth}
    
    \textbf{Answer to RQ2.} \par
     If LLMs focus on detecting single vulnerability, they can achieve a noticeable decrease at false positive rate.

    \end{minipage}
}
\end{center}

\begin{table*}[h!]
    \centering
    \caption{Evaluation via LLMs on multiple Solidity versions}
    \label{tab:result}
    
    \begin{tabular}{ c |c|c|cccc| cccc}
        \toprule
      \multicolumn{1}{c}{\textbf{Version}} & \multicolumn{1}{c}{\textbf{Models}} & \multicolumn{1}{c}{\textbf{Vulnerability} }  & \textbf{TP} & \textbf{FP} & \textbf{TN} &\textbf{FN}   & \textbf{Recall} & \textbf{FPR} & \textbf{Accuracy} & \textbf{Precision}   \\ 
        \midrule
        
     \multirow{30}{*}{\textbf{Solidity v0.8}} & \multirow{6}{*}{ GPT-4o-mini}   & Access Control   & 12 & 205 & 370 & 2 & 0.86 & 0.37 & 0.65 & 0.06\\ 
      & & DOS & 7 & 48 & 385 & 6 & 0.54 & 0.11 & 0.88 & 0.13\\  
       & & Arithmetic Issue
        & \cellcolor{gray!25} 3 & 32 & 446 & \cellcolor{gray!25} 12 & 0.20 & 0.07 & 0.91 & 0.09\\ 
      & & Manipulate Price & 1 & 31 & 435 & 10 & 0.20 & 0.07 & 0.91 & 0.03\\ 
      & & Re-entrancy & \cellcolor{red!25} 2 & 17 & 185 & \cellcolor{red!25} 13 & 0.13 & 0.08 & 0.86 & 0.06\\ 
      & & Oracle Issue & 5 & 30 & 399 & 6 & 0.45 & 0.07 & 0.92 & 0.14\\

       \cmidrule{2-11}
     
      & \multirow{6}{*}{Gemini 1.5 Pro}   & Access Control
     & 13 & 371 & 204 & 1 & 0.92 & 0.65 & 0.37 & 0.03\\ 
     &  & DOS & 12 & 228 & 205 & 1 & 0.92 & 0.53 & 0.49 & 0.05\\ 
      & & Arithmetic Issue  & 14 & 193 & 285 & 1 & 0.93 & 0.40 & 0.61 & 0.07\\ 
      & & Manipulate Price & 10 & 213 & 251 & 0 & 1 & 0.46 & 0.55 & 0.04\\ 
      & & Re-entrancy & 14 & 76 & 126 & 1 & 0.93 & 0.38 & 0.66 & 0.16\\ 
     &  & Oracle Issue & 11 & 195 & 234 & 0 & 1 & 0.45 & 0.56 & 0.05\\

    \cmidrule{2-11}

       & \multirow{6}{*}{Claude 3.5 Sonnet}   & Access Control
         & 9 & 221 & 355 & 4 & 0.69 & 0.38 & 0.62 & 0.04\\ 
      & & DOS & 13 & 299 & 134 & 0 & 1 & 0.69 & 0.33 & 0.04\\ 
      & & Arithmetic Issue  & \cellcolor{gray!25} 3 & 32 & 446 & \cellcolor{gray!25} 12 & 0.20 & 0.07 & 0.91 & 0.09\\ 
      & & Manipulate Price & 7 & 144 & 322 & 4 & 0.64 & 0.31 & 0.69 & 0.05\\ 
       & & Re-entrancy & \cellcolor{red!25} 3 & 23 & 179 & \cellcolor{red!25} 12 & 0.20 & 0.11 & 0.84 & 0.12\\ 
       & & Oracle Issue & 7 & 31 & 398 & 4 & 0.64 & 0.07 & 0.92 & 0.18\\

    \cmidrule{2-11}

       & \multirow{6}{*}{Yi-large}   & Access Control
         & 1 & 18 & 557 & 13 & 0.07 & 0.03 & 0.95 & 0.05\\ 
      & & DOS & 0 & 0 & 433 & 13 & 0 & 0 & 0.97 & 0\\ 
      & & Arithmetic Issue  &\cellcolor{gray!25} 0 & 2 & 476 & 15 & 0 & 0.004 & 0.97 & 0\\ 
      & & Manipulate Price & 0 & 0 & 466 & 11 & 0 & 0 & 0.97 & 0\\ 
      & & Re-entrancy & \cellcolor{red!25} 0 & 0 & 202 & 15 & 0 & 0 & 0.93 & 0\\ 
      & & Oracle Issue & 0 & 0 & 429 & 11 & 0 & 0 & 0.98 & 0\\

    \cmidrule{2-11}
    
       & \multirow{6}{*}{Qwen-plus}   & Access Control
         & 4 & 102 & 473 & 10 & 0.29 & 0.18 & 0.81 & 0.04\\ 
       & & DOS & 1 & 6 & 427 & 12 & 0.08 & 0.01 & 0.96 & 0.14\\ 
       & & Arithmetic Issue  & \cellcolor{gray!25} 1 & 2 & 476 & 14 & 0.07 & 0.004 & 0.97 & 0.33\\ 
      & & Manipulate Price & 1 & 12 & 454 & 10 & 0.08 & 0.03 & 0.95 & 0.08\\ 
      & & Re-entrancy & \cellcolor{red!25} 0 & 7 & 195 & 15 & 0 & 0.03 & 0.90 & 0\\ 
       & & Oracle Issue & 9 & 14 & 415 & 2 & 0.82 & 0.03 & 0.96 & 0.82\\
       
        \bottomrule

        \toprule
      &  \multicolumn{1}{c}{\textbf{Models} } & \multicolumn{1}{c}{\textbf{Vulnerability} }  & \textbf{TP} & \textbf{FP} & \textbf{TN} &\textbf{FN}   & \textbf{Recall} & \textbf{FPR} & \textbf{Accuracy} & \textbf{Precision}   \\ 
        \midrule
        
     \multirow{12}{*}{\textbf{Solidity v0.4}} &  \multirow{2}{*}{ GPT-4o-mini} 
       & Arithmetic Issue
        & 9 & 31 & 5 & 3 & 0.75 & 0.86 & 0.29 & 0.23\\  
     &  & Re-entrancy & 11 & 25 & 11 & 1 & 0.92 & 0.69 & 0.46 & 0.31\\ 

     \cmidrule{2-11}
     
     &  \multirow{2}{*}{Gemini 1.5 Pro}   
       & Arithmetic Issue  & 12 & 36 & 0 & 0 & 1 & 1 & 0.25 & 0.25\\ 
      & & Re-entrancy & 12 & 32 & 4 & 0 & 1 & 0.89 & 0.33 & 0.27\\ 

    \cmidrule{2-11}

      &  \multirow{2}{*}{Claude 3.5 Sonnet} 
       & Arithmetic Issue  & 10 & 29 & 7 & 2 & 0.83 & 0.81 & 0.35 & 0.26\\ 
      & & Re-entrancy & 12 & 19 & 17 & 0 & 1 & 0.53 & 0.60 & 0.39\\ 

    \cmidrule{2-11}

       & \multirow{2}{*}{Yi-large}
       & Arithmetic Issue  &10 & 17 & 19 & 2 & 0.83 & 0.47 & 0.60 & 0.37\\  
       & & Re-entrancy & 8 & 21 & 15 & 4 & 0.67 & 0.58 & 0.48 & 0.28\\ 

    \cmidrule{2-11}
    
       & \multirow{2}{*}{Qwen-plus} 
       & Arithmetic Issue  & 1 & 7 & 29 & 11 & 0.08 & 0.19 & 0.63 & 0.13\\ 
       & & Re-entrancy & 10 & 21 & 15 & 2 & 0.83 & 0.58 & 0.52 & 0.32\\ 
       
        \bottomrule
    \end{tabular}

\end{table*}

\subsection{RQ3: LLM Limitations?}

After evaluating a model's recall rate for different vulnerabilities, we can clearly see that its performance varies when detecting different vulnerabilities. To illustrate, GPT-4o-mini successfully detected almost all ``Access Control" vulnerabilities, with its recall rate reaching over 85\%, while it failed to identify vulnerabilities related to ``Re-entrancy" and ``Arithmetic Issues". \par
Through comparison, we can clearly see that, among these vulnerabilities, detecting the Re-entrancy and Arithmetic Issue vulnerabilities is challenging for general LLMs. The related numbers are highlighted in red and gray in \textbf{Table~\ref{tab:result}}.

\smallskip
\subsubsection{\underline{Arithmetic Issue}}
In order to further analyze the problem in detail, we repeated the experiment and asked models not only to evaluate smart contracts with binary classification, but also generate a related report. \par
\smallskip
\begin{center}
   %\fbox{
   \colorbox{blue!10}{
         \begin{minipage}{0.89\linewidth}
            You are a smart contract auditor. ~\\
            Please review the following smart contract in detail carefully.\par Please confirm whether the smart contract project has “\textbf{A}” vulnerability, and only answer yes or no. If no, please answer with the main reason.[\textbf{content}] 
         \end{minipage}
      }
  % }
\end{center}
\par
\smallskip
In the Arithmetic Issue report, it indicates that the main reason for reporting no vulnerabilities is that the program was written in Solidity v0.8, and the Solidity compiler automatically checks for Overflow/Underflow.\par
\smallskip
\begin{center}
   %\fbox{
   \colorbox{red!15}{
         \begin{minipage}{0.89\linewidth}
            No. This smart contract does not appear to have an Arithmetic Vulnerability. The contract is using Solidity v0.8.2, which includes built-in overflow and underflow checks for arithmetic operations.
         \end{minipage}
      }
   %}
\end{center}
\smallskip

However, ``Arithmetic Issues'' include Overflow/Underflow and incorrect calculations. The models tend to focus heavily on detecting Overflow/Underflow while neglecting calculation-related issues. As a result, if any vulnerability is related to incorrect calculations, the model would possibly fail to detect.

On the contrary, Chen et al.~\cite{chen2023chatgpt} detected Arithmetic Issues with ChatGPT on smart contract dataset smartbugs-curated\cite{durieux2020empirical} , most of which were written in Solidity v0.4. Their results indicate that ChatGPT can easily detect these kinds of issues. The main factor contributing to this is that the Solidity v0.4 compiler doesn't automatically check for Overflow/Underflow. Hence, if there is no library like Safemath to prevent Overflow/Underflow, LLMs would tend to report an Arithmetic Issue vulnerability in the smart contract.

\smallskip
\subsubsection{\underline{Re-entrancy}}

By evaluating the model’s reports for Re-entrancy vulnerabilities, we identified three reasons that may lead to a low recall rate. The main factor is reliance on protective libraries, followed by incorrect checks for the checks-effects-interactions pattern and the complex designs of Re-entrancy attacks.\par

\smallskip
\begin{center}
   %\fbox{
   \colorbox{red!15}{
         \begin{minipage}{0.89\linewidth}
            No. This smart contract is using OpenZeppelin's TimelockController, a well-audited implementation that inherently protects against re-entrancy vulnerabilities.
         \end{minipage}
      }
   %}
\end{center}

During the evaluation for Re-entrancy, models will generally check whether the contract uses protective libraries first, such as ReentrancyGuard a library developed by Openzeppelin to prevent Re-entrancy\cite{Openzeppelin}\cite{chittoda2019mastering}. When this kind of library or framework is detected, models tend to mark the contracts as ``no Re-entrancy". The smart contracts written in Solidity v0.8, including those we used in previous experiment, have all used protection mechanism. \par
However, it can't guarantee that the smart contract won't be vulnerable to a Re-entrancy attack. In other words, there is still a high possibility of a Re-entrancy attack if the smart contract's protection mechanism is weak, let alone some tricky attack that can bypass the protection mechanism.

To validate our hypothesis, we conducted an additional experiment to comprehensively evaluate the performance of LLMs in detecting ``Arithmetic Issue" and ``Re-entrancy" vulnerabilities in smart contracts written in Solidity v0.4. \par 
In the experiment, after reviewing the data, we decided not to use the dataset smartbugs-curate~\cite{durieux2020empirical} discussed earlier. The main reason is that the smart contracts are all very short, 
averaging 23 lines of code. This would significantly reduce the LLM's error rate and cannot reflect the true situation. 

Therefore, we used the dataset Messi-Q Smart-Contract-Dataset~\cite{liu2023rethinking}. Published in 2023, the programs are long enough and cover both types of vulnerabilities. This allows us to conduct a thorough and meticulous comparative experiment.

\smallskip
We collected 48 smart contracts from the dataset for testing, of which 12 were labeled with specific vulnerabilities that we aimed to test, including Arithmetic Issue and Re-entrancy.\par 
Using the same LLMs and the same prompts, we conducted the comparative experiment, and the results are presented in \textbf{Table~\ref{tab:result}}. The findings indicate that general LLMs can identify most of ``Arithmetic Issue" and ``Re-entrancy" vulnerabilities with a high recall rate, thus validating our hypothesis.

\begin{figure}[h]
\centering
\subfigure[False Positive Rate]{\label{fig1}        
\includegraphics[width=0.95\linewidth]{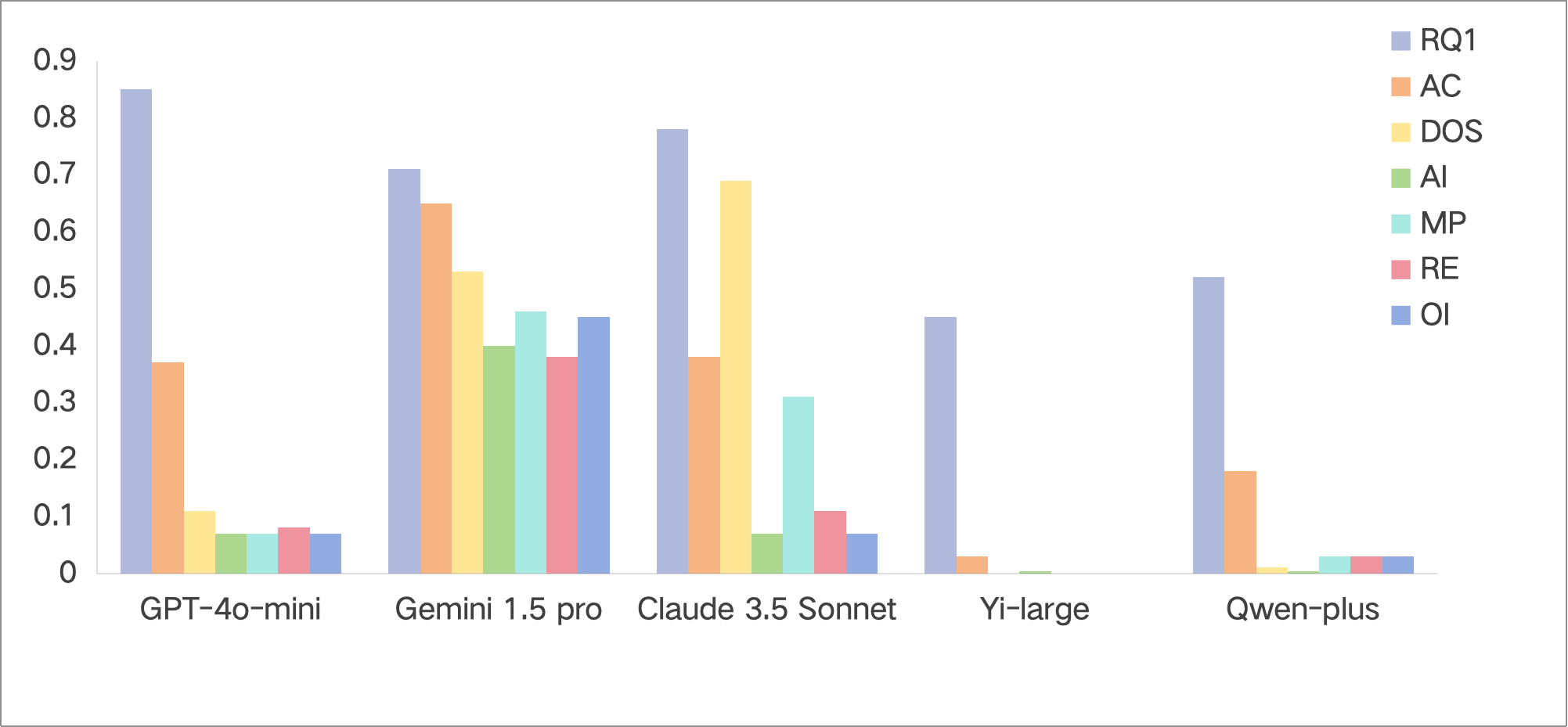}
%\caption{Comparison of False Positive Rate}
}
\subfigure[RE Recall Rate]{\label{fig2}
\includegraphics[width=0.95\linewidth]{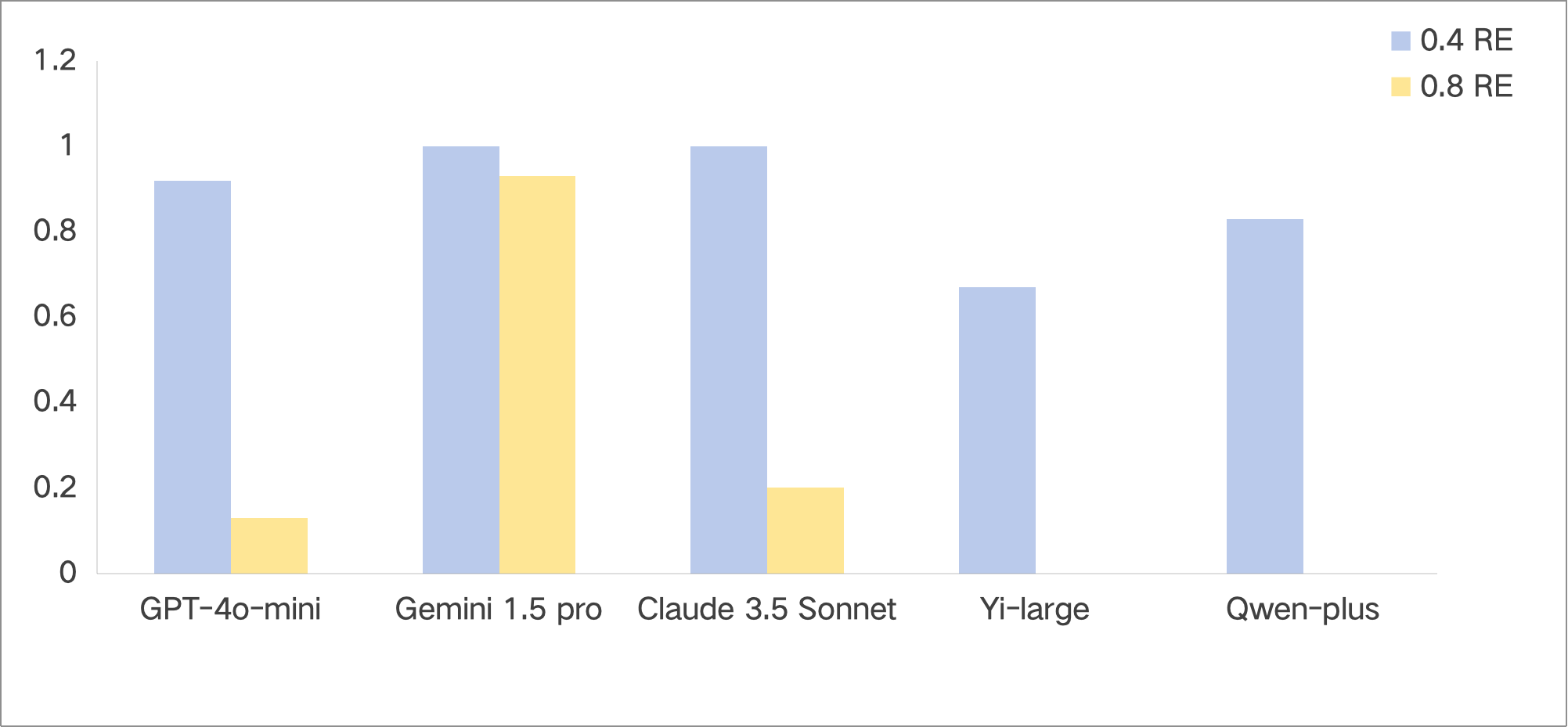}
%\caption{Comparison of RE Recall Rate}
}
\subfigure[AI Recall Rate]{\label{fig3}
\includegraphics[width=0.95\linewidth]{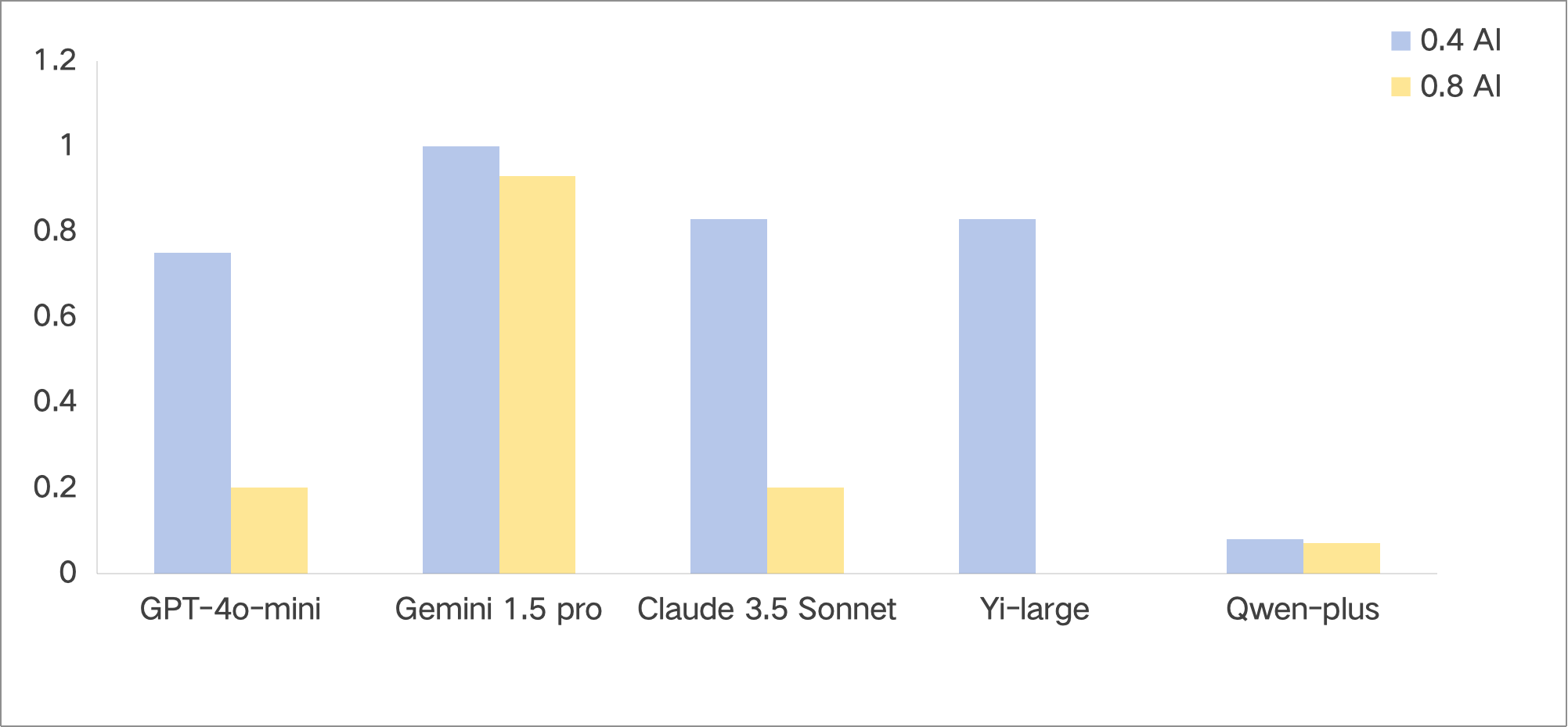}
}
\caption{Comparisons}
\label{fig-comp}

 \vspace{-0.2in}
\end{figure}

The comparative result is shown in \textbf{Fig.~\ref{fig2}} and \textbf{Fig.~\ref{fig3}}. It reveals that smart contracts are highly likely to be marked with a Re-entrancy vulnerability if no protective libraries or frameworks are used in the program. At the same time, general LLMs may fail to identify Re-entrancy vulnerabilities if the smart contract uses a protective mechanism, which is the usual case in smart contracts written in Solidity v0.8.

\subsubsection{\underline{Model's Difference}}

In the evaluation, after redesign of prompt, even all the model's false positive rate generally decreased, the models still showed different levels of false positive rate. The striking difference is the two models (Yi-large and Qwen-plus) maintained a extraordinary low false positive rate and recall rate, which means they discover limited vulnerabilities, but also did not incorrectly report too many vulnerabilities,  while Gemini 1.5 Pro hunted much more faults and maintained a extremely high recall rate. But it also shows a highest false positive rate. In other words, if a model has a stronger ability to detect vulnerabilities, it would tend to have a higher false positive rate.

\begin{center}
\fbox{%
    \begin{minipage}{0.98\linewidth}

    \textbf{Answer to RQ3.} \par
     LLMs face new challenges in detecting vulnerabilities in smart contracts written in Solidity v0.8, particularly those related to Arithmetic Issues and Re-entrnacy. At the same time, high detection capability may lead to a higher false positive rate. 
 
    \end{minipage}
}
\end{center}

%================================================
\section{Related Work}
\label{sec-rw}
%================================================
\subsection{LLMs in Program Analysis}
Commercial LLMs, including OpenAI’s GPT-4o, Google’s Gemini, Anthropic's Claude, have gained rapid development in recent years, and enthusiastically promoted as tools to help programmers in coding tasks like detecting bugs and code generation\cite{sandoval2023lost}. LLMs could be particularly useful to help developers with their cybersecurity
needs, as humans typically produce and miss many security
relevant bugs. This issue was highlighted in the 2022 GitLab
Survey, noting that ``developers do not find enough bugs early enough” and ``do not prioritize the bug remediation” when developing\cite{ullah2024llms}.

Numerous works have contributed to evaluating the performance of LLM in coding related challenges: vulnerable code
generation\cite{pearce2022asleep}, code repair\cite{pearce2023examining}, detecting vulnerabilities in the coding\cite{thapa2022transformer}, and a detailed evaluating framework\cite{ullah2024llms}.

\subsection{Evaluating Smart Contract Security}

\begin{table*}[h]
 \caption{Smart Contract Detection with LLM }\label{tab-stateofart}
 \label{node}
  \centering
  \begin{threeparttable}
    \resizebox{\linewidth}{!}{  
    \begin{tabular}[t]{c|ccccc}
    \toprule
     \multicolumn{1}{c}{{\textbf{ }} } &  \textit{\textbf{Dataset}} &  \textit{\textbf{Solidity}} & \multicolumn{1}{c}{\textbf{\textit{LLM Mdoel}}}   & \multicolumn{1}{c}{\textbf{\textit{Prompting method}}}  & \multicolumn{1}{c}{\textbf{\textit{Results}}}   \\  
     
     \midrule
     
    \cite{chen2023chatgpt}~(2023)  & smartbugs-curated & v0.4 & GPT-3.5, GPT-4, GPT-4o  & Detect 9 vulnerability types &  A high recall rate but limited precision   \\  

   \cite{david2023you}~(2023)   & 52 smart contracts & unknown   & GPT-4-32k, Claude 1.3 & Binary classification & Identify
vulnerability types in 40\% of cases \\

    \cite{luo2024fellmvp}~(2024) & \makecell{Messi-Q} & v0.4 & Multiple Fine-tuned models & Detect one specific fault & Achieve 98.8\% accuracy and 88\% F1 score\\ 

    \midrule

    \textbf{Ours}  & \makecell{TOP200, Web3Bugs, \\Messi-Q} & v0.4/0.8 & \makecell{GPT-4o-mini, Claude 3.5, \\Yi-large,  Gemini 1.5 Pro, \\Qwen plus} & \makecell{Binary classification \\ Detect one specific fault} & \makecell{v0.4: High recall rate and high FPR \\ v0.8 Low recall rate and low FPR} \\  

    \bottomrule
    \end{tabular}
    }
      \begin{tablenotes}
       \footnotesize
       \item[] \scriptsize \textbf{Abbreviation}: \textbf{Messi-Q} is short for Messi-Q Smart Contract Dataset.
  \end{tablenotes}
      \end{threeparttable}
      \vspace{-0.2in}
\end{table*}

Auditing smart contract is a time-consuming and challenging assignment. Aside from manual assessment by a human smart contract auditor, there are numerous tools and techniques have been designed to perform security analyses of smart contract. Based on their attributes, tools can be classified as static analysis, symbolic execution, buzzing, and machine learning\cite{sendner2024large}, while JF. Ferreira et al. proposed a large framework, including more than 19 supported smart contract scanners in 2020\cite{ferreira2020smartbugs} and 2023\cite{di2023smartbugs} respectively, and C. Sendner et al. conducted a large scale research on these tools\cite{sendner2024large}. Apart from it, after many efforts to automatically discover vulnerabilities of computer programs with LLM, it has been discussed that LLM may have the potential to reveal vulnerabilities of smart contracts\cite{david2023you}.

\subsubsection{\underline{LLMs}}
David et al.~\cite{david2023you} first introduced LLMs in evaluating smart contracts, and conducted experiments on a limited dataset with LLM models GPT-4-32k and Claude-v1.3-100k. In the same year, C. Chen et al.~\cite{chen2023chatgpt} analyzed the performance of GPT-4 series models in detecting vulnerabilities of smart contracts, including GPT-3.5, GPT-4, and GPT-4o. They found that ChatGPT has different detection performance for different vulnerabilities, with a relatively high recall rate but low precision rate. However, the datasets they used in the experiment are smartbugs-curated~\cite{durieux2020empirical}, which is slightly outdated and most of smart contracts were written in Solidity v0.4, while the current version of smart contracts is mainly Solidity v0.8. There has been a huge advancement in Solidity-related libraries. 

To improve LLM's performance in detecting vulnerabilities of smart contracts, several researchers~\cite{ince2024detect}\cite{ma2024combining} begin to fine-tune LLMs. There are also new smart contract scanners combining LLM and classical technologies, including static analysis \cite{sun2024gptscan} and fuzzing test~\cite{shou2024llm4fuzz}.

\subsubsection{\underline{Static Analysis}}
Static analysis, also called static code analysis, is a method of computer program debugging that is done by examining the code without executing the program~\cite{w7}. Tikhomirov et al. proposed SmartCheck in 2018. It translates Solidity source code into an XML-based intermediate representation and checks it against XPath patterns\cite{tikhomirov2018smartcheck}. In the same year, Grishchenko et al. presented the first sound and automated static analysis for EVM bytecode, which is practical and scales to large contracts\cite{grishchenko2018ethertrust}. Then Brent et al. presented Vandal, a security analysis framework consists of an analysis pipeline that converts low-level Ethereum Virtual Machine (EVM) bytecode to semantic logic relations\cite{brent2018vandal}. It successfully analyzed over 95\% of all 141k unique contracts with an average runtime of 4.15 seconds. In 2019, Feist et al. designed Slither, the first static open source extended-able framework for detecting vulnerabilities of smart contracts. It was written in Python3 and supports detecting more than 80 types of vulnerabilities~\cite{feist2019slither}.

\subsubsection{\underline{Symbolic Execution}}
Symbolic execution of programs is supplying symbols representing arbitrary values, instead of supplying the normal inputs to a program (e.g. numbers)\cite{king1976symbolic}. In 2016, L. Luu build a symbolic execution tool called Oyente to find potential security bugs. Among 19, 336 existing Ethereum contracts, Oyente flags 8,833 of them as vulnerable\cite{luu2016making}, but it is only able to dicover 4 types of vulnerabilities. In 2018, Torres et al. introduced Osiris -- a framework that combines symbolic execution and taint analysis, in order to accurately find integer bugs in Ethereum smart contracts.\cite{torres2018osiris}
In the same year, Nikolić et al. implemented Maian, the first tool for specifying and reasoning about trace properties, which employs inter-procedural symbolic analysis and a concrete validator for exhibiting real exploits\cite{nikolic2018finding}. They reproduced real exploits at a true positive rate of 89\%, yielding exploits for 3,686 contracts. In 2020, Frank et al. designed and implemented ETHBMC. They also performed a large-scale analysis of roughly 2.2 million accounts currently active on the blockchain and automatically generated 5,905 valid inputs, which trigger a vulnerability\cite{frank2020ethbmc}.

\subsubsection{\underline{Fuzzing}}
Fuzzing or Fuzz testing is an automated software testing technique that involves providing invalid, unexpected, or random data as inputs to a computer program. The program is then monitored for exceptions such as crashes, failing built-in code assertions, or potential memory leaks\cite{miller1995fuzz}. In 1988, Miller et al. proposed Fuzzing Test first\cite{miller1990empirical}. He et al. introduced fuzzing in detecting vulnerabilities in smart contracts in 2019\cite{he2019learning}. They proposed to learn an effective and fast fuzzer from symbolic execution, by phrasing the learning task in the framework of imitation learning. In 2020, Nguyen proposed sFuzz, which was applied to more than 4 thousand smart contracts\cite{nguyen2020sfuzz}. Choi et al. performed a lightweight dynamic data-flow analysis to collect data-flow-based feedback to effectively guide fuzzing in 2021. They implemented ideas on a practical open-source fuzzer named SMARTIAN\cite{choi2021smartian}. SMARTIAN can discover bugs in real-world smart contracts without the need for the source code.

\subsubsection{\underline{Machine Learning}}
Machine learning (ML) is a branch of artificial intelligence (AI) focused on enabling computers and machines to imitate the way that humans learn, to perform tasks autonomously, and to improve their performance and accuracy through experience and exposure to more data.\cite{ML} In 2019, J. W.Liao et al. Proposed SoliAudit. Combined machine learning and fuzz tesing, It was able to detect vulnerabilities without expert knowledge or predefined patterns\cite{liao2019soliaudit}. They also created a gray-box fuzz testing mechanism, which consists of a fuzzer contract and a simulated blockchain environment for on-line transaction verification. In 2023, C. Sender et al. first applied deep transfer learning in detecting vulnerabilities\cite{sendner2023smarter}. They also demonstrated that it achieves an average F1 score of 98\% on six vulnerability types in initial training and yields an average F1 score of 96\% in the transfer learning phase on five additional vulnerability types.

%================================================
\section{Concluding Remarks}
\label{sec-conclusion}
%================================================
In this paper, we build upon existing research on detecting vulnerabilities in (up-to-date) smart contracts using LLMs via a series of well-designed experiments. %We formulated three research questions to explore their potential and limitations. 

For RQ1, to initially test the false positive rate of current mainstream LLMs, we conducted an experiment on a widely used smart contract dataset. The results revealed that the false positive rate of general LLMs is typically high, exceeding 70\%. For RQ2, we successfully reduced the false positive rate of general LLMs with an optimized and reasonable prompt, and validated this through a comprehensive experiment on a well-audited dataset primarily containing smart contracts written in Solidity v0.8, which is prevalently used at present. The results also indicated that the detection abilities of LLMs vary, and highlighted the challenges they face in detecting certain specific vulnerabilities. For RQ3, we carefully designed a comparative experiment between the two versions of code and analyzed the root causes of the limitations during detection. Our findings reveal the reliance of LLMs on identifying changes in newly introduced libraries and frameworks during detection. This opens up new approaches for enhancing the performance of LLMs in identifying vulnerabilities.

%We plan to extend our work along two critical pathways. The first is to enhance our evaluation framework, enabling it to uncover more limitations in the detection capabilities of LLMs. Furthermore, building on our evaluation results, we aim to optimize model performance by fine-tuning the models and incorporating other program analysis techniques, such as Symbolic Execution, Fuzzing, and Machine Learning.

%================================================
{\footnotesize \bibliographystyle{IEEEtran}
\bibliography{bib}}
%================================================

\end{document}